# Hydrogen Peroxide electrosynthesis: A comparative study employing Vulcan Carbon modification by different $MnO_2$ nanostructures

*João Paulo C. Moura*[1], *Vanessa S. Antonin*[1], *Aline Trench*[1], *Mauro C. Santos*[*1]

1) *Laboratório de Eletroquímica e Materiais Nanoestruturados, Centro de Ciências Naturais e Humanas, Universidade Federal do ABC, CEP 09210-170, Rua Santa Adélia 166, Bairro Bangu, Santo André, SP, Brazil*

*Corresponding Author:*

**E-mail:* mauro.santos@ufabc.edu.br

**Abstract**

The electrochemical performances of the α-$MnO_2$/Vulcan XC-72 and δ-$MnO_2$/Vulcan XC-72 nanostructures in hydrogen peroxide ($H_2O_2$) electrosynthesis were compared herein. Both materials were synthesized by a simple hydrothermal route. Their structures and morphologies were analyzed by SEM, HRTEM, XPS, Raman Scattering and XRD, and their ORR electrochemical properties and $H_2O_2$ electrosynthesis efficacies were investigated in alkaline NaOH solutions applying the rotating ring-disk electrode (RRDE) technique. Gas diffusion electrode (GDE) setups in acid media aiming at $H_2O_2$ formation were also performed. The 3% δ-$MnO_2$/C and 1% α-$MnO_2$/C electrocatalysts were more efficient and selective than pure Vulcan XC-72 through the ORR 2-electron pathway in the RRDE essays. Concerning $H_2O_2$ electrogeneration using GDE, the 1% α-$MnO_2$/C electrocatalyst displayed better activity, with peroxide accumulation of 402.6 mg/L at -1.9 V (vs Ag/AgCl) after 120 min, 48 % higher than pure Vulcan XC-72 GDE. These results can be ascribed to a synergistic effect between α-$MnO_2$ and Vulcan XC-72, as well as oxygen functional acid species improvement, increasing electrocatalytic surface hydrophilicity and enhancing $H_2O_2$ electrosynthesis.

**Keywords:** $MnO_2$ nanostructures, Vulcan Carbon, Hydrogen Peroxide Electrosynthesis, Oxygen Reduction Reaction.

## 1. Introduction

The Oxygen Reduction Reaction (ORR) is a key electrochemical process of many environmental-friendly chemical and electrochemical processes, presenting high applicability and, thus, being exceedingly investigated and explored in various scientific fields [1]. The ORR is a multi-electron process that, depending on the mechanism reaction, can be of vital use in several fields. The four-electron pathway of this reaction, for example, is desirable for energy applications, applied to fuel cells [2,3] and metal-air batteries [4,5] The two-electron pathway, on the other hand, is eco-friendly, economic and attractive for $H_2O_2$ electrogeneration, drastically reducing the use of reagents required in conventional synthesis methods, applied in organic solvent-free conditions, also allowing *in-situ* $H_2O_2$ generation, facilitating product transport, storage, and handling [6–8].

Efficient $H_2O_2$ production via the ORR requires effective and selective catalysts, leading to many investigations to find ideal compounds[9], and numerous materials, such as noble metals [10], transitional metals [11], oxide nanoparticles [12] and metal-free carbon-based materials [13] have been suggested as alternative electrocatalysts for the ORR two-electron pathway. Manganese dioxide ($MnO_2$) has attracted increasing attention in this regard, due its abundance, low cost and environmental friendliness, comprising a promising catalyst for energy storage/conversion devices, such as batteries and supercapacitors, also exhibiting electrochemical and catalytic properties, of interest in oxygen electroreduction processes[14,15]. Crystallographic structures and morphology, *i.e.*, particle shape and size, specific surface areas, electronic structures, manganese valence and defects, play a crucial role in the chemical properties of $MnO_2$ nanostructures [16]. These features can be controlled by synthetic methods. Selvakumar, et al.[17] , for example, synthesized and compared five types of $\alpha$-$MnO_2$ nanostructures, reporting that morphology significantly influences the number of electrons transferred in the ORR, altering the pathway from 2 to 4 electrons. In another study,

Cheng [18] and Xiao [12] reported both morphology influences and the dependence of ORR catalysis on crystallographic structures, indicating high ORR potential. These studies also investigated bulk $MnO_2$ as an electrocatalyst, reporting poor hydrogen peroxide ($H_2O_2$) selectivity as an ORR pathway product.

Common carbon materials can be combined with oxide materials aiming at electrochemical-based applications, such as conductivity enhancement and electrocatalyst surface-property improvements [19,20]. Previous studies by our research group have indicated that ORR $2e^-$ selectivity can be influenced by the loading oxide-carbon catalyst [21–24]. Pinheiro et. al.[22] demonstrated promising results for the ORR $2e^-$ pathway using a $CeO_2$ nanorod/Vulcan XC-72 (1% loading w/w), electrocatalyst, reporting a synergistic effect between $CeO_2$ and Vulcan XC-72, improving catalyst efficiency and selectivity, with greater oxide-carbon loadings leading to decreased ORR selectivity.

In this context, this study aimed to evaluate the synthesis and characterization of different nanostructured crystalline $MnO_2$ phases synthesized by a simple hydrothermal method, and compare their performance in modifying Vulcan XC72 carbon EDGs for the ORR $2e^-$ pathway and *in-situ* $H_2O_2$ electrosynthesis.

## 2. Experimental

### 2.1. Materials

Potassium permanganate ($KMnO_4$), manganese sulfate monohydrate ($MnSO_4 \cdot H_2O$) and Polytetrafluoroethylene (PTFE) were purchased from Sigma-Aldrich. All the chemical reagents were of analytical grade and used without additional purification treatment.

### 2.2. $MnO_2$ nanostructure synthesis

The $MnO_2$ materials were synthesized by a hydrothermal method. Briefly, 0.4 g of $MnSO_4 \cdot H_2O$ and 1.0 g of $KMnO_4$ were mixed in 30 mL deionized water, transferred to a 100 mL Teflon-lined stainless-steel autoclave and heated at 140 °C for different times (12 h and 24 h) in a furnace. Following the hydrothermal treatment, the autoclave was cooled down to room temperature and the obtained products were washed with water and ethanol several times and dried in an oven at 70 °C for 6 h.

### 2.3. $MnO_2$/C electrocatalyst preparation

The $MnO_2$ electrocatalysts supported on carbon Vulcan XC-72 (1, 3, 5 and 10 % w/w) were prepared by a wet impregnation method[25]. An appropriate amount of $MnO_2$ was added to 0.5 g of Vulcan XC-72, suspended in 30 mL of deionized water in an ultrasonic bath for 30 min and placed under vigorous magnetic stirring for 2 h. The synthesized electrocatalysts were then dried in an electric oven at 80 °C.

## 2.4. Electrocatalyst characterization

X-ray powder diffraction (XRD) was performed employing a D8 Focus diffractometer (Bruker AXS) operating at 40 kV and 40 mA with CuKa radiation in the 10°– 90° 2θ range at a scan rate of 2° $min^{-1}$. Raman scattering was applied using a Horiba-Jobin-Yvon, model T64000 coupled to a laser light at 532 nm and applying an exposure time of 40 s. Material morphologies and microstructures were investigated using a FESEM JEOL JSM-7401 operating at 30 kV and a HRTEM JEOL JEM 2100 microscope operating at 200 kV. XPS analyses were performed on a Scienta Omicron ESCA + spectrometer (Germany) using monochromatic Al Kα (1486.7 eV). The inelastic background of C 1s high-resolution corelevel spectra was subtracted using Shirley's method. The spectra were fitted without placing constraints using multiple Voigt profiles. The width at half maximum (FWHM) varied between 1.0 and 2.4 eV.

Wettability was characterized through electrocatalyst contact angle measurements using a goniometer (GBX Digidrop). Briefly, 1 mg/mL dispersions (catalyst powder/deionized water) were prepared through sonication for 1 min using a tip ultrasonicator. A 40 µL aliquot of each dispersion was then deposited on a glass plate and dried to form a thin and homogeneous film. Subsequently, 10 µL of deionized water were dropped on the film surface to determine the contact angle. Measurements were obtained in triplicate using the Windrop software.

## 2.5. Oxygen Reduction Reaction Study

Electrochemical measurements were performed employing an Autolab PGSTAT 302 N Potentiostat/galvanostat with a rotating ring-disc electrode system (RRDE) (Pine Instruments) using a 125 mL electrochemical cell. A 2 $cm^2$ platinum counter electrode (Pt), a reference

Hg/HgO electrode containing a 4 mol $L^{-1}$ NaOH solution and a working electrode composed by a GC disc (area = 0.2475 $cm^2$) and gold (Au) ring (area = 0.1866 $cm^2$) with an experimental collection factor of N = 0.28 and theoretical of N = 0.22, and 1 mol $L^{-1}$ NaOH (Synth) was used as the support electrolyte for the electrochemical measurements. The electrocatalysts were deposited on the glassy carbon disc working electrode as thin, porous layers by drop casting. Dispersions containing 1 mg/mL of the electrocatalyst in water were homogenized using a tip ultrasonicator. After homogenization, 20 µL of the dispersions were deposited in the disc electrode surface. After drying, 20 µL of a 1:100 Nafion solution (v/v, Nafion: deionized water) were pipetted on the film covering the working electrode and dried. The NaOH 1 mol $L^{-1}$ electrolyte was previously saturated with oxygen for 30 min for all electrochemical analyses, maintaining the same flow throughout al electrochemical measurements, performed in duplicate at a scan rate of 5 mV $s^{-1}$ at room temperature.

### 2.6. ***In situ*** **$H_2O_2$ electrosynthesis study**

*In situ* $H_2O_2$ electrosynthesis was accomplished using a gas diffusion electrode (GDE) cathode prepared by a hot-pressing procedure using $MnO_2$/Vulcan XC-72 or Vulcan XC-72 containing 20 % (m/m) of a 60 % aqueous PTFE dispersion, according to the method described previously by our research group [23]. Electrogeneration was carried out in an undivided 350 mL electrochemical cell using a reference Ag/AgCl electrode, a Pt counter electrode (5 $cm^2$) and the fabricated GDE with continuous $O_2$ supplying at 0.2 bar. The electrolyte comprised 0.1 mol $L^{-1}$ $H_2SO_4$ and 0.1 mol $L^{-1}$ $K_2SO_4$. The electrolysis process was performed for 2 h at -0.7, -1.3, -1.9, -2.5 V *vs* Ag/AgCl. The amount of peroxide was determined by UV-Vis spectrophotometry using $(NH_4)_6Mo_7O_{24}\cdot 4H_2O$ ($2.4 \times 10^{-3}$ mol $L^{-1}$) in $H_2SO_4$ (0.5 mol $L^{-1}$), as described previously [23,26].

## 3. Results and discussion

3.1. **$MnO_2$ nanostructure characterizations**

The XRD patterns of the $MnO_2$ materials synthesized under different hydrothermal treatments times are displayed in Fig 1. Hydrothermal times directly affected the crystalline $MnO_2$ material structure, with the product prepared for 12 h identified as a pure birnessite phase ($\delta$-$MnO_2$) according to the Inorganic Crystal Structure Database (ICSD) card no. 9001270 (monoclinic structure with a space group C2/m). A typical morphology is observed for $\delta$-$MnO_2$ by FESEM, depicted in Fig. 2(a), indicating a nanoflower-like morphology with several intersected nanosheets. Higher hydrothermal treatment times led to the complete transition of $\delta$-$MnO_2$ to the $\alpha$-$MnO_2$ crystalline phase. The XRD pattern of the 24 h hydrothermally-treated product is indicated in Fig 1, corresponding to the $\alpha$- $MnO_2$ phase according to ICSD card no. 20227 (tetragonal structure with space group I4/m). An altered morphology is noted with the crystalline phase conversion. Figure 2(b) indicates needle-like nanorods structures in the $\alpha$-$MnO_2$ phase, indicating that hydrothermal treatment time is key in controlling the crystalline phase and morphologic structure of the compound during $MnO_2$ synthesis.

**<Insert Figure 1>**

**<Insert Figure 2>**

The $\alpha$-$MnO_2$ nanostructure was examined by High Resolution (HRTEM), depicted in Fig. 3. The HRTEM image and the corresponding Fast Fourier Transform (FFT) of the selected area indicates a d spacing of about 2.38 Â from the lattice-resolved TEM image, consistent with the spacing between (211) planes and with the powder XRD pattern.

**<Insert Figure 3>**

Following synthesis, the obtained $MnO_2$ nanostructures were supported on Vulcan XC-72 carbon by wet impregnation to prepare the $MnO_2$/C electrocatalysts at 1, 3, 5 and 10 % w/w loadings. Figure 4 depicts the SEM images of the δ-$MnO_2$/Vulcan XC-72 (Fig. 4a and 4b) and α-$MnO_2$ /Vulcan XC-72 (Fig. 4c and 4d) materials, demonstrating that the wet impregnation method provides good nanostructure distribution onto the carbon support.

**<Insert Figure 4>**

Raman spectroscopy allows for the assessment of the structural properties of carbonaceous materials, applied herein to assess potential defects and disorders in the prepared $MnO_2$/Vulcan XC-72-based electrocatalysts. Carbon Black materials like Vulcan XC-72 display two quite sharp modes, the *D* peak around 1350 $cm^{-1}$ and the *G* peak around 1580–1600 $cm^{-1}$, usually assigned to zone center $E_{2g}$ symmetry phonons and *K*-point $A_{1g}$ symmetry phonons, respectively[27].

The G peak occurs due to the vibrational modes present in all C=C bonds, and is active in all $sp^2$ sites. The G peak corresponds to the peak generated by first order Raman scattering involving only one phonon. The D peak, although also related to sp2 bonds, refers to the breathing-type vibrational modem that becomes active only in the presence of structural disorders. Thus, the intensity ratio between peak D and peak G ($I_D/I_G$) is often used to assess the disorder degree of carbon networks [28,29].

The calculated $I_D/I_G$ ratios are listed in Fig 5(a). The highest ratio was observed for the α-$MnO_2$/C electrocatalyst compared to the δ-$MnO_2$/C electrocatalyst and the pure Vulcan XC-72 material, while the lowest ratio was observed for δ-$MnO_2$/C. These findings indicate that α-$MnO_2$/C electrocatalysts exhibit higher defect densities compared to the other analyzed electrocatalysts This amorphous carbon material trend may be associated to oxygen-loading functional groups and indicates increased oxygen vacancies, which may enhance electron transfer and, thus, ORR activity [3,30,31].

The contact angle measurements used to investigate electrocatalyst surface hydrophilicity are displayed in Fig 5(b). The highest contact angle of 23.5° was observed for pure Vulcan XC-72 carbon, decreasing with $MnO_2$ nanostructure incorporation, reaching 12.4° and 14.4° for δ-$MnO_2$, and α-$MnO_2$ respectively. Contact angle and hydrophilicity are, therefore, correlated to one another, as the greater the contact angle, the less hydrophilic is the material, and the lower the contact angle, the more hydrophilic is the material[32]. In this regard, the carbon support modification was shown to improve material hydrophilicity properties. Electrocatalyst wettability, on the other hand, is closely related to dissolved-$O_2$ transfer efficiency, improving $O_2$ transport and active site adsorption, consequently increasing ORR activity[33,34].

XPS analyses were conducted to assess the surface chemical state of the synthesized electrocatalysts. The deconvoluted C 1s core-level high-resolution spectrum for Vulcan XC-72 and α-$MnO_2$/C is shown in Fig 5(c). The C 1s spectra exhibit similar peak shapes for both electrocatalysts, with peaks observed at 284.1 eV and 285.0 eV, corresponding to the aromatic (C-C) and aliphatic (C-H) phases, respectively. Additionally, peaks related to oxygen-containing functional groups such as C—O, C═O, and O═C—O, as well as plasmon transitions, are observed at 285.9 eV, 286.6 eV, 288.3 eV, and 289.6 eV, respectively [35–37].

The quantitative fitted XPS analysis revealed an increase in local oxygen content, with 17.7 at. % in Vulcan and 20.0 at. % in α-$MnO_2$-modified Vulcan. This confirms a higher presence of oxygenated species on the catalysts' surface, which contributes to the hydrophilic surface character. This, in turn, improves the mass transfer process between dissolved oxygen and the catalyst surface. Additionally, the presence of oxygen-containing groups enhances the electron donor ability of the carbon in the local vicinity [38,39].

**<Insert Figure 5>**

3.2. **Electrochemical measurements**

Electrochemical ORR activity and electrocatalyst selectivity analyses were carried out by linear sweep voltammetry (LSV) using a rotating ring-disk electrode (RRDE) using a $O_2$-saturated 1 mol $L^{-1}$ NaOH solution. Figures 6a and 6b, indicate the steady-state polarization ring and disk curves, respectively. Reference materials for the 2-electron and 4-electron ORR mechanisms, comprising Vulcan XC-72 and Pt/C, respectively, were examined as controls and comparisons. In the LSV technique, the $O_2$-saturated solution is electro-reduced in the disc covered by the electrocatalyst and the electrogenerated $H_2O_2$ is carried to the ring by rotation and there undergoes oxidation. The higher the ring oxidation currents, the higher the electrocatalyst ORR selectivity for $H_2O_2$ electrogeneration.

The obtained polarization curves indicate a similar profile for the Vulcan XC-72 and $MnO_2$/Vulcan XC-72-based electrocatalysts, proving that both follow the 2-electron ORR pathway. The best $MnO_2$/C-based electrocatalyst for each of the crystalline phases were 3% δ-$MnO_2$ and 1% α-$MnO_2$, at ring currents of 58 μA and 41 μA, respectively, both higher than the determined Vulcan XC-72 ring currents (35 μA), suggesting that $MnO_2$ modification increases $H_2O_2$ electrogeneration selectivity. The results can be used to deduce ORR reaction kinetics via the Koutecky-Levich equation (Eq.1), as follows:

$$\frac{1}{i} = \frac{1}{i_k} + \frac{1}{i_d} = \frac{1}{nFAkC_{O_2}} - \frac{1}{0.62nFADo_2^{2/3}\nu^{-1}C_{O_2}\omega^{1/2}} \quad (1)$$

Where $i_k$ and *id* represent the kinetics and diffusion limiting current, respectively; *n* is the electron number involved in the ORR; *F* is the faraday constant (96485 C $mol^{-1}$), *A* is the area effectively covered with the catalyst; *k* is the electron transfer rate for the ORR (m $s^{-1}$), $C_{O_2}$ is the $O_2$ saturation concentration in solution, $Do_2^{2/3}$is the $O_2$ diffusion coefficient in solution, *v* is the kinematic viscosity and *ω* is the angular rotational speed of the electrode (rpm).

**<Insert Figure 6>**

The K-L plot is depicted in Fig 7a. The effective number of electrons transferred in the ORR is calculated through the slope of the K-L plot. The K-L plot is given in a straight line of I vs $\omega^{-1/2}$ whose slope value can be associated to the number of electrons involved in the electrochemical process at the same potential. The Vulcan XC-72 Carbon-based and Pt catalysts are reference materials for the 2-electron and 4-electron ORR pathways, respectively. The linear plots slopes (Figure 7a) are similar for Vulcan XC-72 and for $MnO_2$/C-based electrocatalysts and different from the curve profile of the Pt/C Alfa Aesar, which represents a 4-electron reduction ORR pathway. These results confirm that the number of electrons involved in the ORR for all $MnO_2$ catalysts applied for Vulcan modification is closer to 2 electrons. Analyses concerning the number of catalyst electrons transferred in the ORR and selectivity toward $H_2O_2$ production were performed based on Equations 2-3:

$$X_{H_2O_2} = \frac{2I_r/N}{-I_d+I_r/N} \quad (2)$$

$$n_t = 2\,[2 - (X_{H_2O_2})] \quad (3)$$

Where $I_r$, $I_d$ and $N$ are the ring current, disk current and collection efficiency of the RRDE, respectively. The results demonstrate that 1% α-$MnO_2$ and 3% δ-$MnO_2$ applied as Vulcan XC-72 catalysts exhibit high selectivity, with $X_{H2O2} > 70\%$, alongside a lower number of electrons involved in the ORR, with significantly improved catalytic ORR activity compared to the pure Vulcan XC-72 reference catalyst. Tafel plots were investigated to better assess the parameters of ORR kinetics, shown in Fig.7b. The Tafel slopes were -61 mV $dec^{-1}$, -63 mV $dec^{-1}$ e 56 mV $dec^{-1}$ for 1% α-$MnO_2$, 3% δ-$MnO_2$ and Vulcan XC-72, respectively. The values of slope for all electrocatalysts in an alkaline medium were around 60 mV $dec^{-1}$, this slope values indicates a fast electron transfer followed by a rate-determining step in the chemical process[40], suggesting that they also follow the Temkin isotherm[41,42]. The slight difference in the Tafel slope between the Vulcan XC-72 and $MnO_2$/C electrocatalysts may be attributed to a change in the nature of the active site (associated with an increase in oxygenated species on the surface) involved in the ORR[43,44]

A specific relationship between the $I_D/I_G$ ratio and the selectivity of $H_2O_2$ as product against oxide loading is shown in Fig 7d. The correlation between Raman characterization and peroxide yield for the crystalline phases follows the same trend with respect to the oxide:carbon loading. For δ-$MnO_2$/C, a trend is observed that follows the volcano type curve, while for α-$MnO_2$/C the same behavior was not observed, and from this relationship shown in the figure, we can conclude that the activity of both electrocatalysts is governed by the amount of surface defects (information provided by the $I_D/I_G$ ratio). For α-$MnO_2$/C the trend does not follow the volcano type curve probably due to the morphology and dimensions of the α-$MnO_2$ nanoparticles. We believe that the morphology of the $MnO_2$ nanorods and their unique distribution and interaction cause the synergistic effect between the two materials to exhibit this unusual trend. These findings provide new insights into synergistic electrocatalytic activity for the design of $MnO_2$/Vulcan XC-72 electrocatalysts. Furthermore, it is important to stress that high loadings of $MnO_2$ does not permit the enhancement of hydrogen peroxide electrogeneration, since in this case the amount of carbon becomes low, and the effect is synergic between the oxides (low amount) and carbon XC-72.

**<Insert Figure 7>**

Thus, starting from more active catalysts assessed in the ORR-RRDE experiments, the promising samples were tested to evaluate the real amount of $H_2O_2$ electrogeneration using a gas diffusion electrode (GDE). Figure 8 displays $H_2O_2$ electrogeneration as a function of electrolysis times at different applied potentials obtained with unmodified Vulcan XC-72 and the modified Vulcan XC-72 ($\alpha$-$MnO_2$/C and $\delta$-$MnO_2$/C) GDEs. The $H_2O_2$ GDE accumulation increased with decreasing potentials (more negative potentials) for all GDEs, which can be associated to a decrease or absence of parallel reactions, such as $H_2O_2$ reduction, $H_2$ evolution and $H_2O$ as a by-product of the ORR pathway, which can take place at higher potentials [45].

**<Insert Figure 8>**

The $H_2O_2$ GDE accumulation employing Vulcan XC-72, $\delta$-$MnO_2$/C and $\alpha$-$MnO_2$/C were 270.3, 203.5, 402.6 mg/L at -1.9 V (vs Ag/AgCl) after 120 min of electrolysis, respectively. The GDE based on the 1% $\alpha$-$MnO_2$/C electrocatalyst exhibited excellent catalytic activity towards $H_2O_2$ electrogeneration in an acid solution containing $H_2O_2$, 48 % and 98 % higher than Vulcan XC-72 and 3% $\delta$-$MnO_2$/C, respectively (including at – 2.5 V). The higher $H_2O_2$ production observed when employing α-$MnO_2$/C may be associated to a synergistic effect between α-$MnO_2$ and Vulcan XC-72. Furthermore, more oxygen functional acid species on the catalyst surface increased hydrophilicity in comparison with the δ-$MnO_2$ phase and pure Vulcan XC-72 [21,23,46], probably related to more specific carbon property modifications due to the presence of the $MnO_2$ nanorods (α-$MnO_2$) on Carbon Vulcan XC-72.

These results are compared with GDEs reported in the literature, as shown in Table 1. Our results achieved excellent activity, in which other works were carried out under similar operating conditions (mainly the maintenance of electrolysis in potentiostatic operation). It can be observed that 1% α-$MnO_2$/C presented optimal $H_2O_2$ accumulation compared to some recent reports. It is important to point out that most works reporting MnO2-based electrocatalysts lead to a 4-electron ORR pathway, which involves the complete reduction of oxygen to water [12,18,47–49]. Furthermore, the activity of the synthesized electrocatalyst was greatly enhanced with a small amount of $MnO_2$ modifying Vulcan XC-72, which is highlighted in this present paper.

**<Insert Table 1>**

## 4. Conclusions

A simple hydrothermal route was applied for the synthesis of two distinct $MnO_2$ nanostructures. In addition to hydrothermal time control, the δ-$MnO_2$ to α-$MnO_2$ nanostructure morphology and crystalline phases can be adapted. Comparisons between structural and electrochemical $MnO_2$/Vulcan XC-72 electrocatalyst properties for ORR were conducted, indicating that the 3% δ-$MnO_2$/C and 1% α-$MnO_2$/C electrocatalysts were more efficient and selective than pure Vulcan XC-72 for $H_2O_2$ electrogeneration in RRDE essays. Concerning $H_2O_2$ GDE electrogeneration, the 1% α-$MnO_2$/C exhibited better catalytic $H_2O_2$ electrogeneration activity, with peroxide accumulation 48 % and 98 % higher than Vulcan XC-72 and 3% δ-$MnO_2$/C, respectively. This may be ascribed to the synergistic effect between α-$MnO_2$ and Vulcan XC-72, and by the effect of improved oxygen functional acid species at the catalyst surface. The α-$MnO_2$/C electrocatalyst is more specific for carbon property modifications towards the ORR $2e^-$ pathway and exhibits better $H_2O_2$ ORR electrosynthesis, indicating its potential application as a promising catalyst in wastewater treatment applying advanced electrochemical oxidation processes.

## Acknowledgments

The authors wish to thanks to the financial support of the following Brazilian Research Funding Institutions: Fundação de Amparo à Pesquisa do Estado de São Paulo (FAPESP, 2021/05364-7, and 2017/10118-0), Coordenação de Aperfeiçoamento de Pessoal de Nível Superior (CAPES, 88887.354751/2019-00) and CNPq (303943/2021-1).

**Figure 1.**

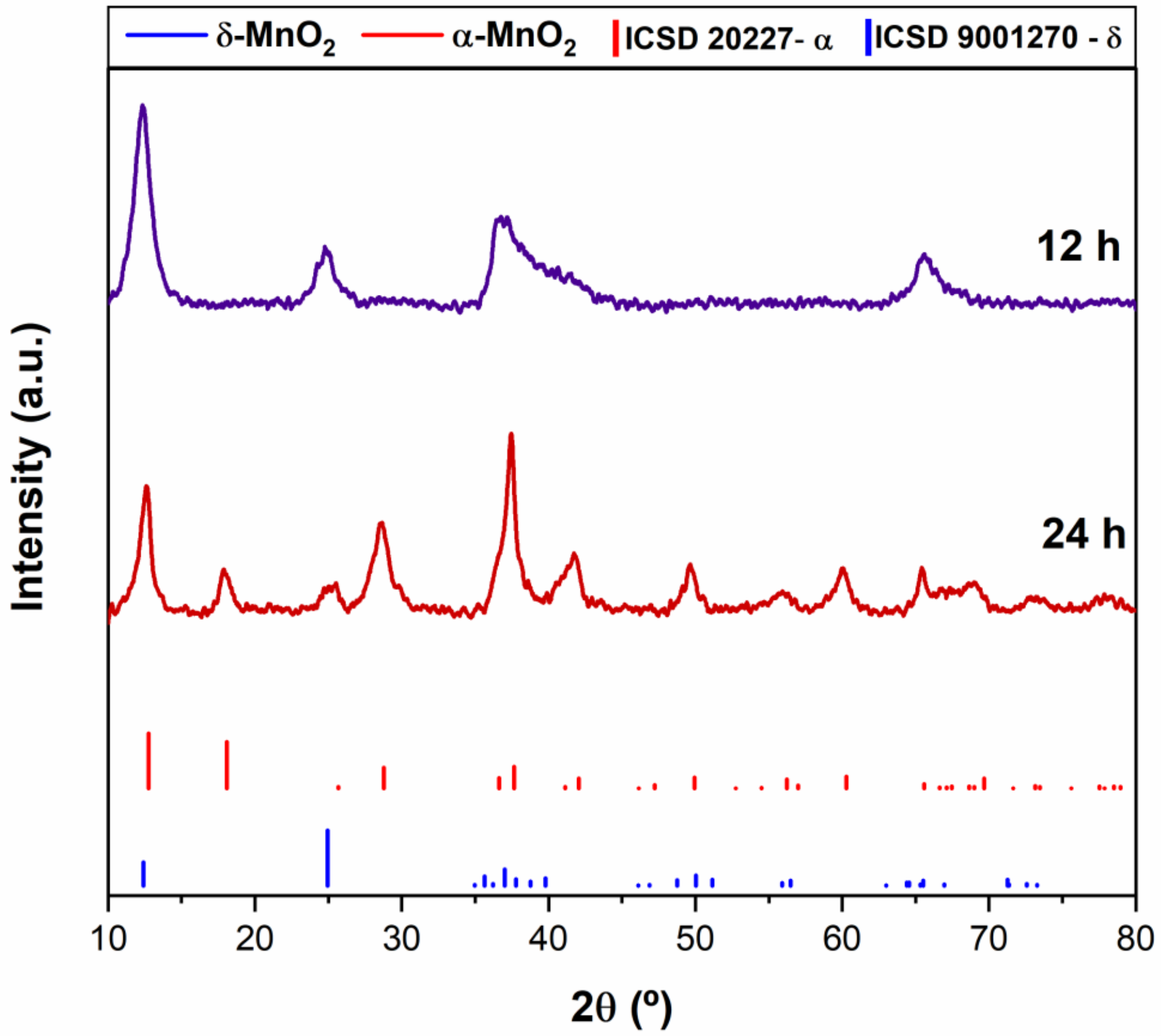

**Figure 1:** XRD patterns of the hydrothermally synthesized $MnO_2$ oxides. The vertical sticks below are relative to the peak positions and relative peak intensities using ICSD cards as reference.

**Figure 2.**

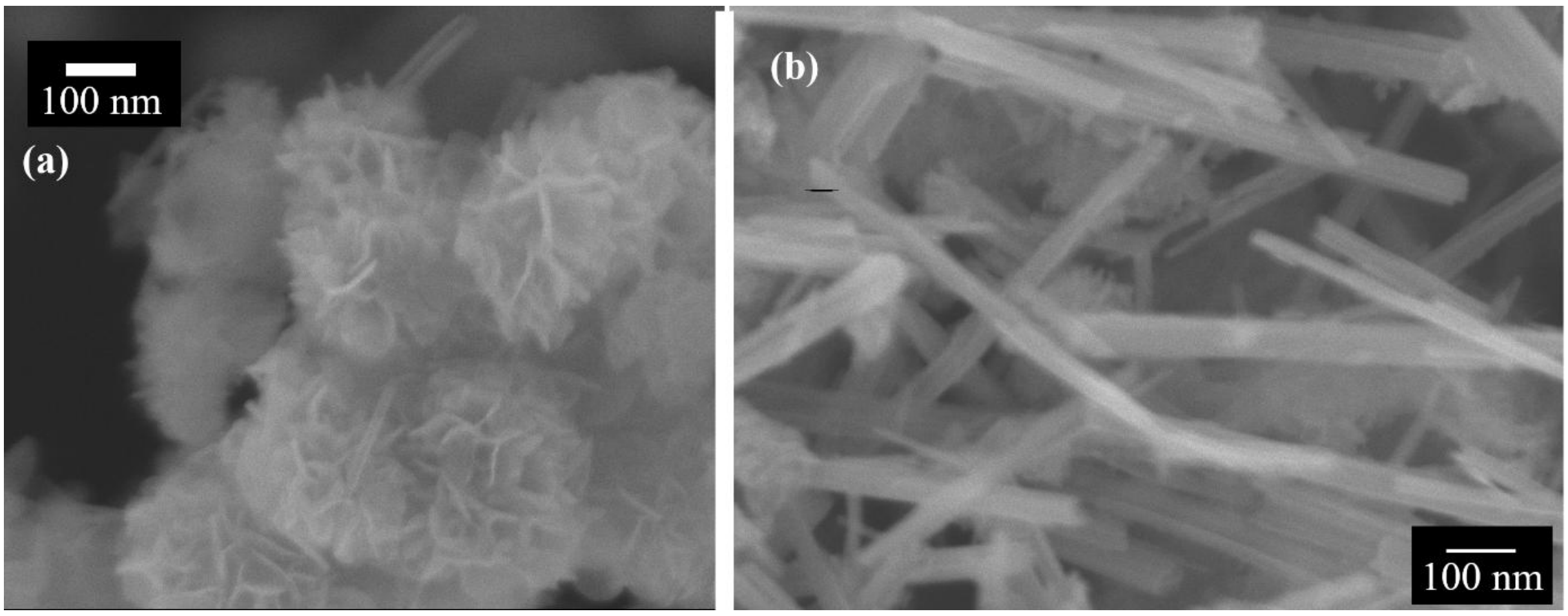

**Figure 2:** SEM images of the hydrothermally synthesized $MnO_2$ oxide at 140 ºC for different hydrothermal times, (a) $\delta$-$MnO_2$ and (b) $\alpha$-$MnO_2$.

**Figure 3.**

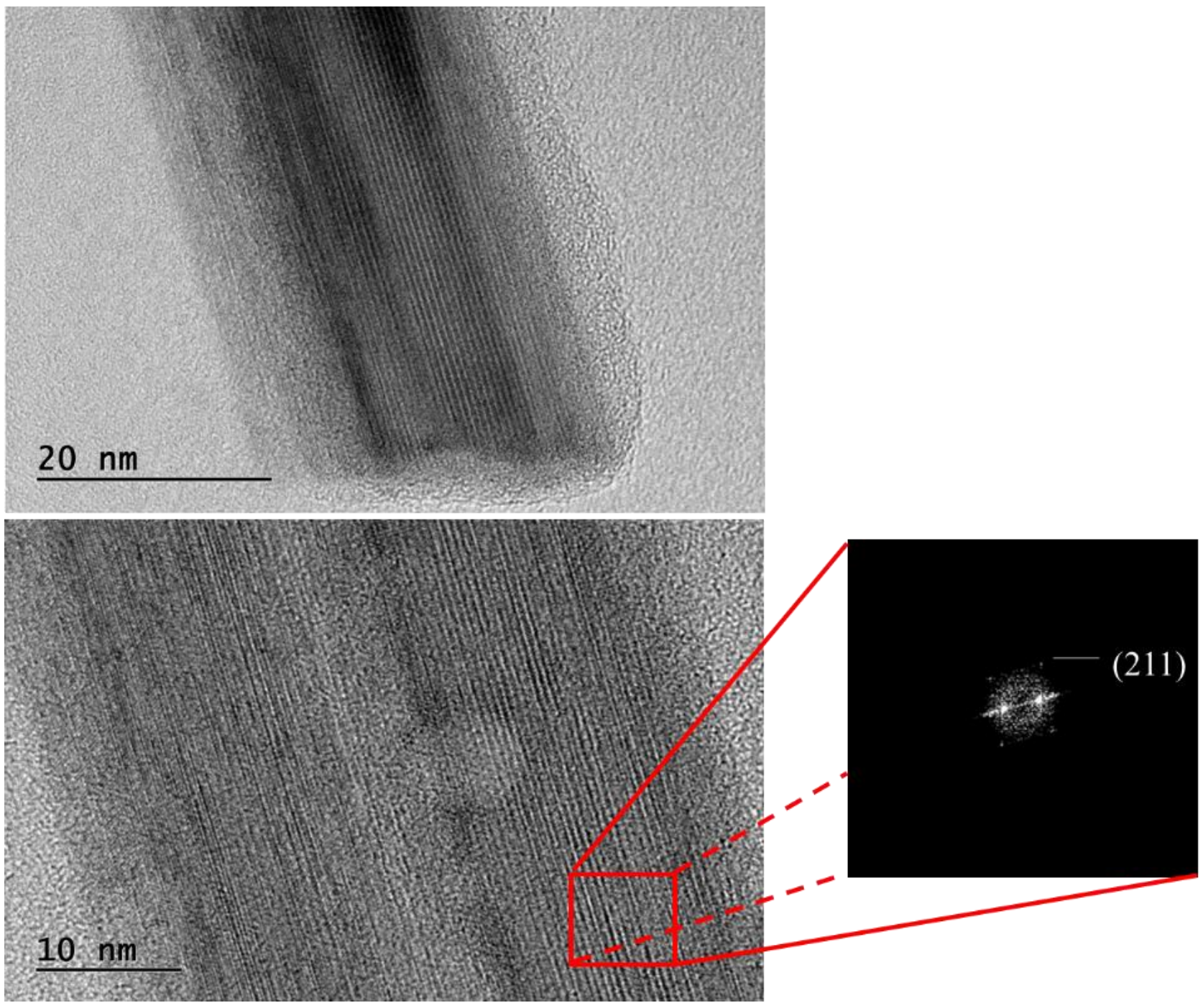

**Figure 3:** HRTEM α-$MnO_2$ image and its corresponding FFT.

**Figure 4.**

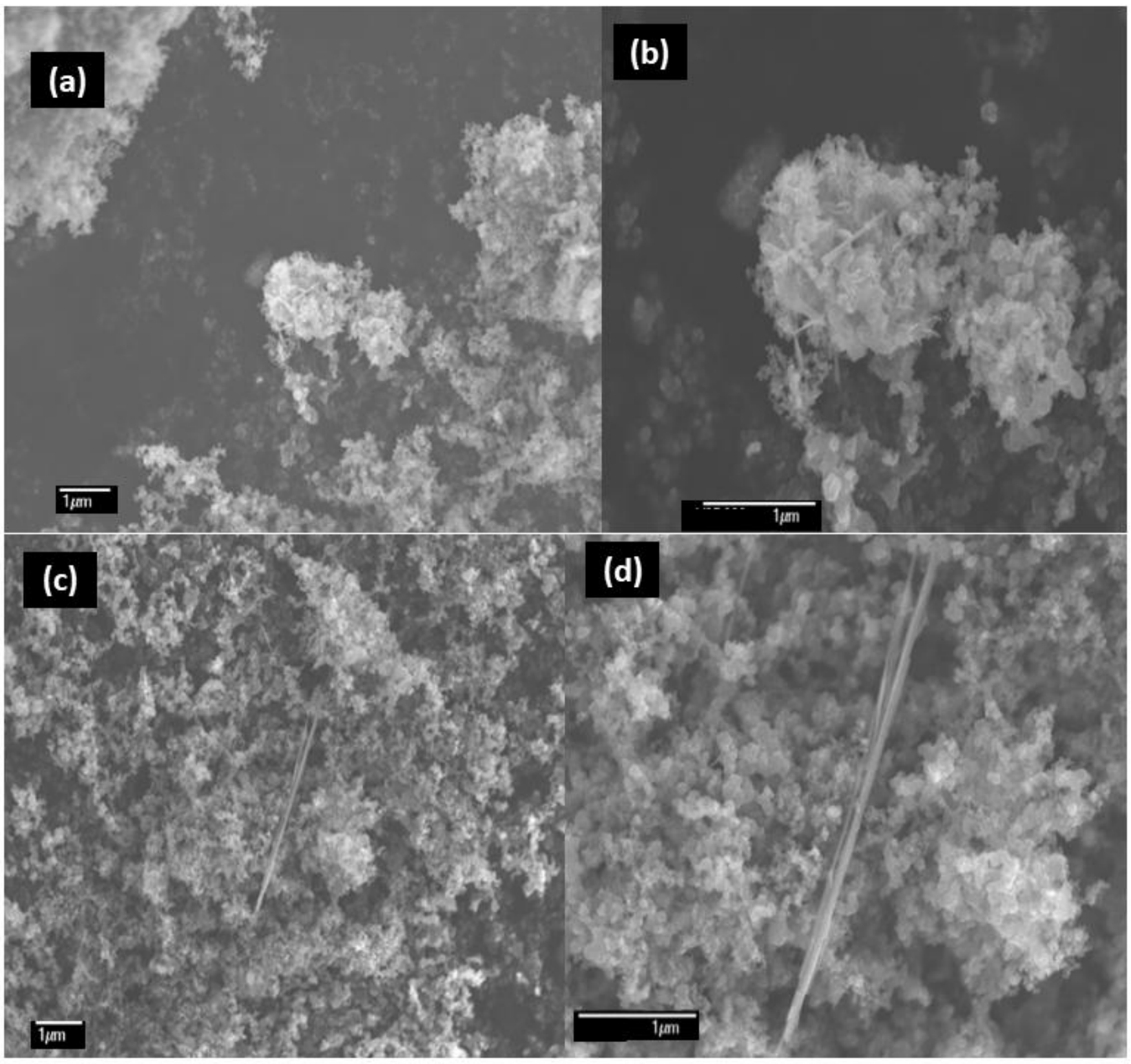

**Figure 4.** (a-b) SEM images of the 3% δ-$MnO_2$/Vulcan XC-72 and (c-d) 1% α-$MnO_2$ /Vulcan XC-72 w/w materials obtained by the wet impregnation method.

**Figure 5.**

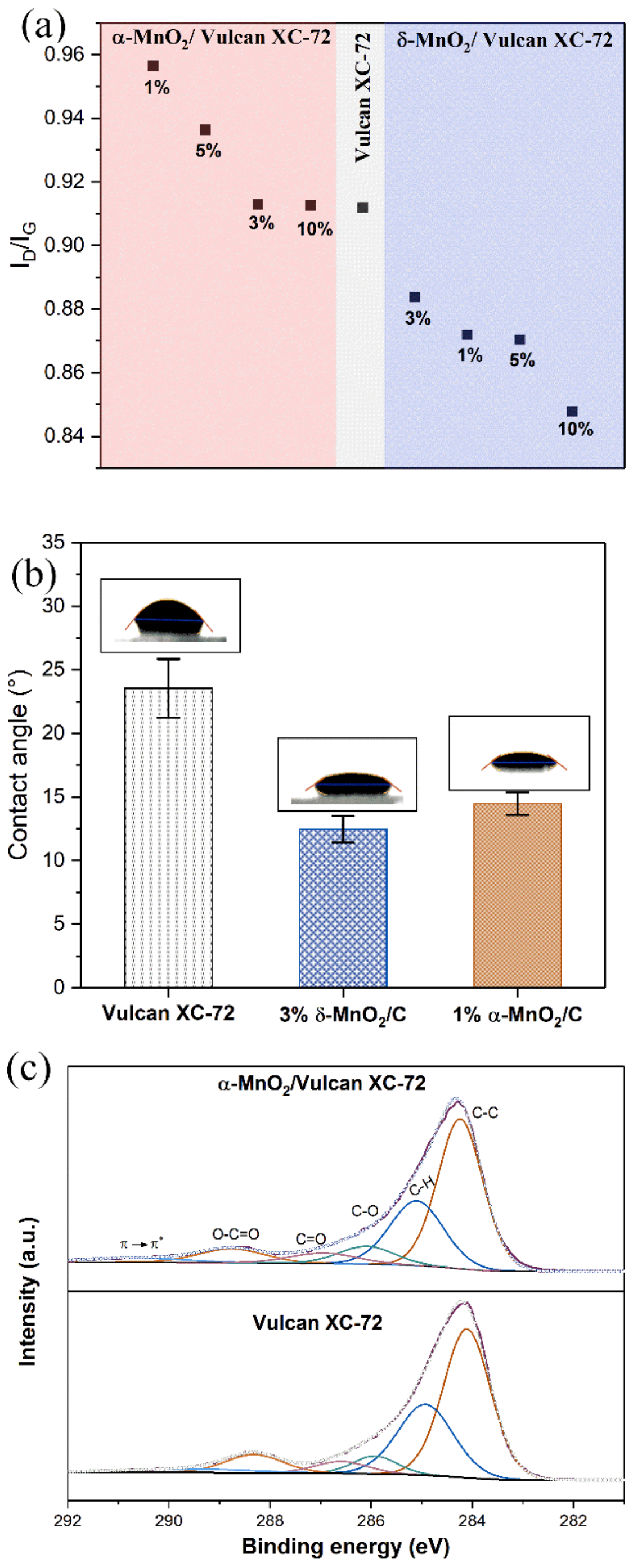

**Figure 5:** (a) $I_D/I_G$ ratios obtained by Raman spectra; (b) Electrocatalyst contact angles. Inset: deionized water droplets on the electrocatalyst layers and (c) Deconvoluted C 1s XPS spectra of Vulcan XC-72 carbon and 1% α-$MnO_2$ /C.

**Figure 6.**

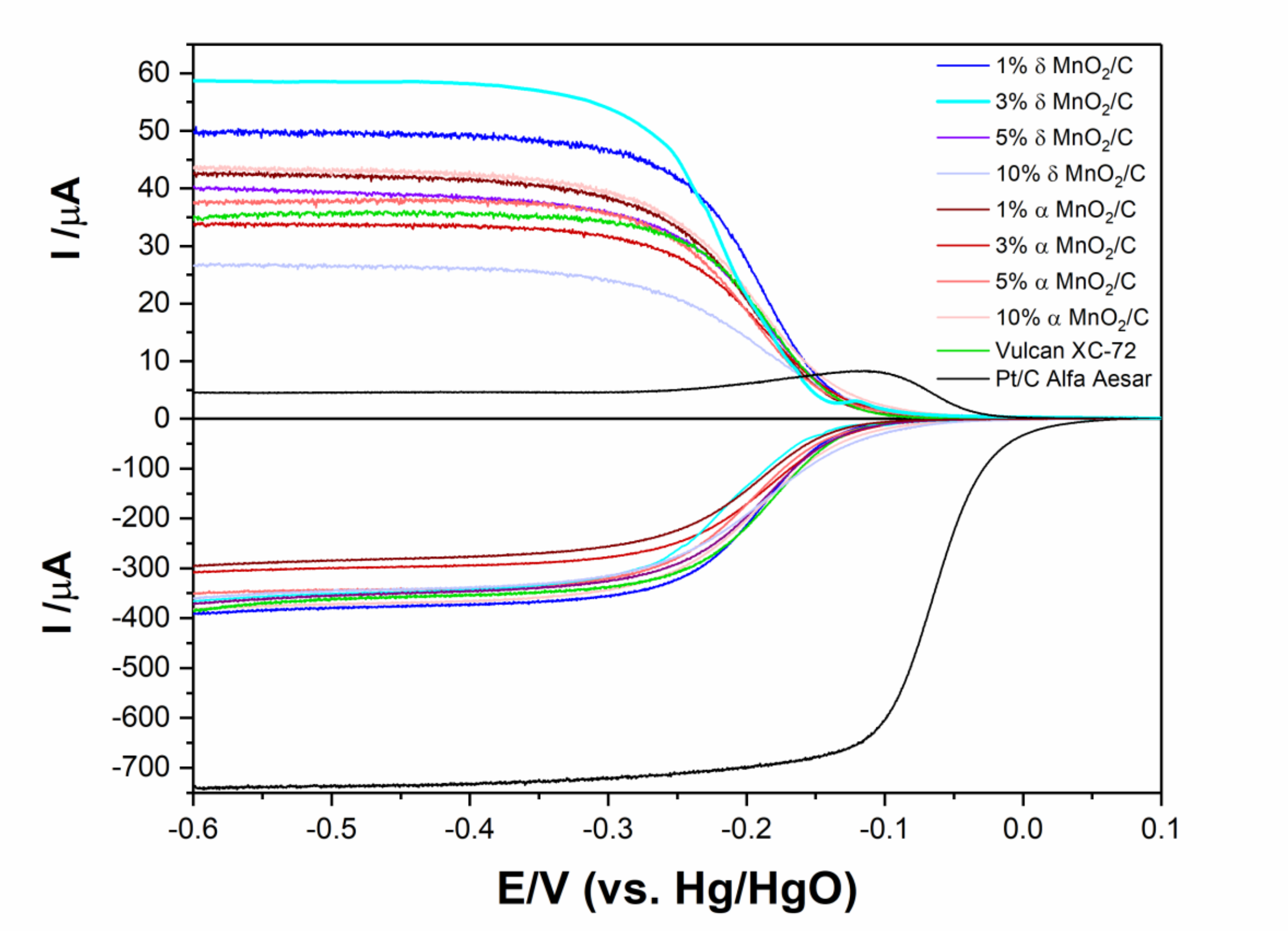

**Figure 6:** Steady state polarization curves for the ORR employing $MnO_2$/C-based electrocatalysts and the reference materials Vulcan XC-72, and Pt/C AA in a $O_2$-saturated NaOH 1 mol $L^{-1}$ solution at a scan rate of 5 mV $s^{-1}$ at 1600 rpm. (a) Ring current $E_{ring}$ = 0.3 V (b) negative disk current sweep.

**Figure 7.**

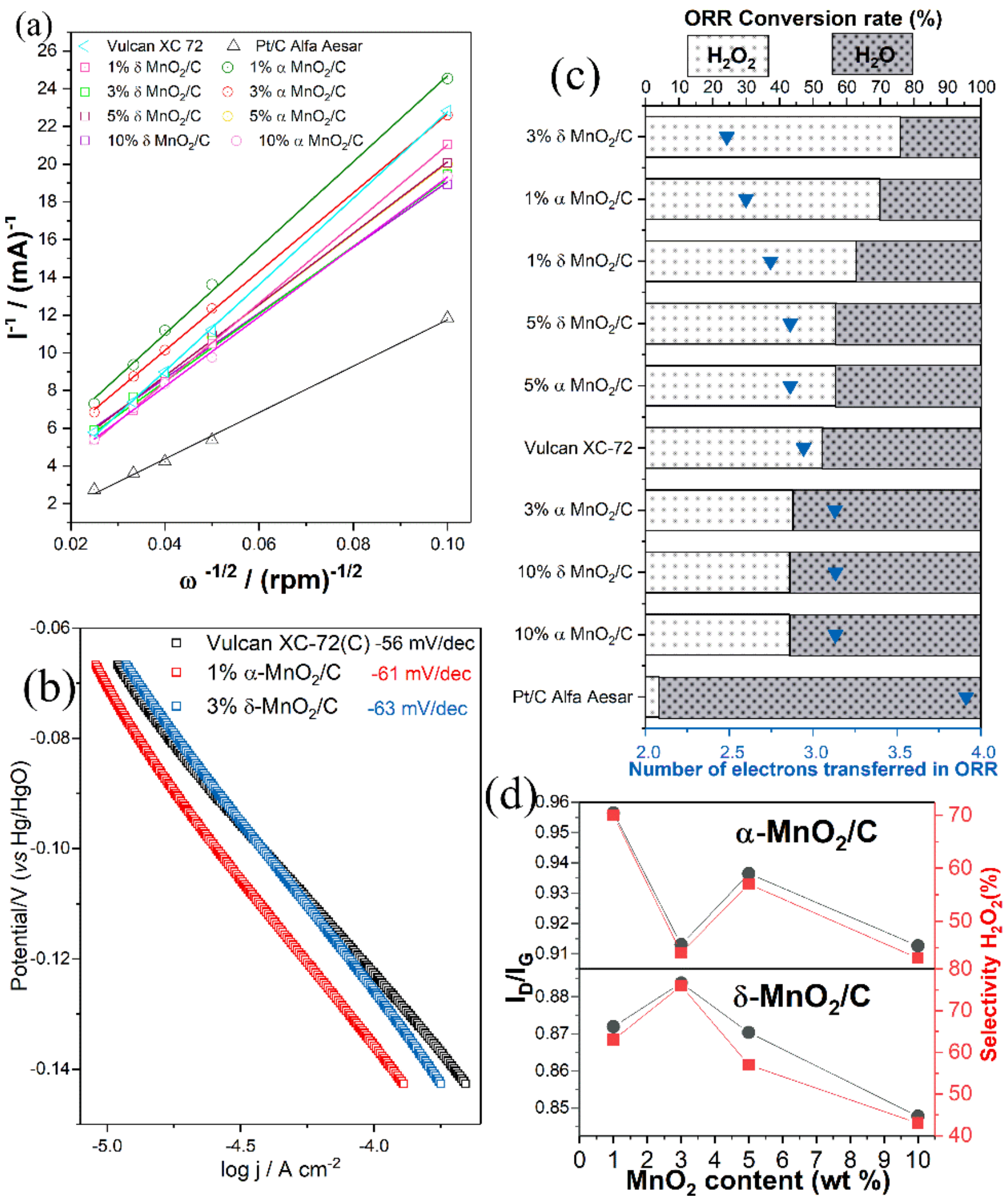

**Figure 7**. (a) ORR K-L plots, (b) Tafel plots, (c) Number of transferred electrons and $O_2$ conversion rate on the $MnO_2$/C-based electrocatalysts, Vulcan XC-72 and Pt/C AA in a 1 mol $L^{-1}$ NaOH $O_2$-saturated solution and (d) relation $I_D/I_G$ *vs* $H_2O_2$ selectivity in function of $MnO_2$ content.

**Figure 8.**

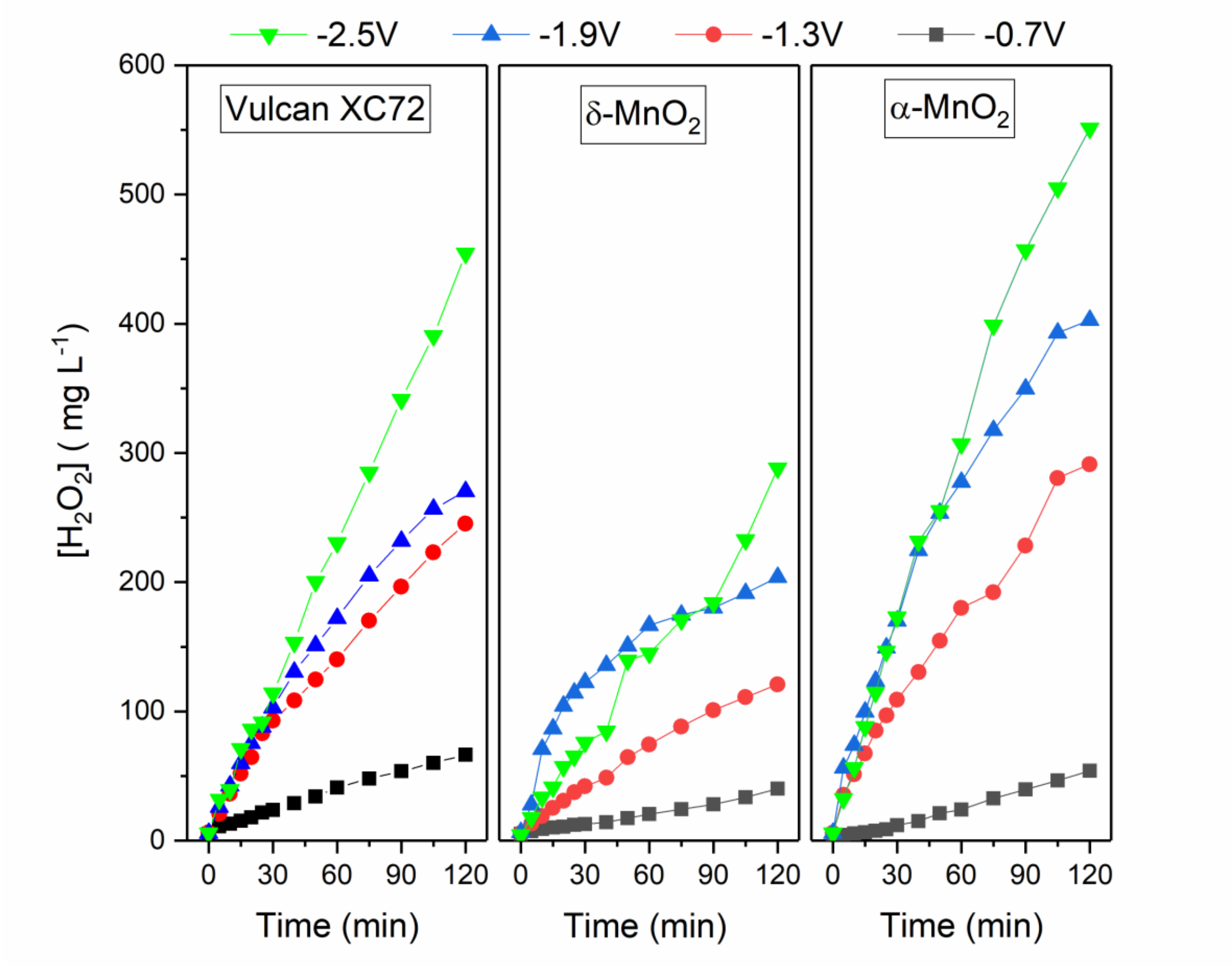

**Figure 8:** $H_2O_2$ electrogeneration at different potentials (*vs* Ag/AgCl) as a function of electrolysis time for different GDEs.

Table 1. Reported results of $H_2O_2$ Electrogeneration using GDEs under potenciostatic.

| **GDE catalyst material** | **Experimental conditions** | **$[H_2O_2]$ accumulated ($mg\ L^{-1}$)** | **Ref.** |
|---|---|---|---|
| 1% α-$MnO_2$/VXC (w/w) | At -1.9 V (vs Ag/Ag/Cl), $H_2SO_4$ 0.1 mol $L^{-1}$ + $K_2SO_4$ 0.1 mol $L^{-1}$ pH 2, 120 min of electrolysis | 402 | This work |
| 3% MnO2/VCX (w/w) | At -1.1 V (vs Ag/Ag/Cl), $H_2SO_4$ 0.1 mol $L^{-1}$ + $K_2SO_4$ 0.1 mol $L^{-1}$ pH 2, 120 min of electrolysis | 391 | [24] |
| 1% W@Au/VCXR (w/w) | At -1.1 V (vs Ag/Ag/Cl), $H_2SO_4$ 0.1 mol $L^{-1}$ + $K_2SO_4$ 0.1 mol $L^{-1}$ pH 2, 120 min of electrolysis | 120 | [23] |
| *5% Co-Porphyrin/PL6 (w/w)* | At -1.75 V (vs Ag/Ag/Cl), $K_2SO_4$ 0.1 mol $L^{-1}$ pH 2.5, 90 min of electrolysis | 340 | [50] |
| *5% $Ta_2O_5$/PL6 (w/w)* | At -1.0 V (vs Ag/Ag/Cl), $K_2SO_4$ 0.1 mol $L^{-1}$ pH 2, 120 min of electrolysis | 27.9 | [51] |

Abbreviations: VXC = Vulcan XC 72; VXCR = Vulcan XC 72R; PL6 = Printex L6